\documentclass[12pt,aps,prd,preprint,tightenlines,superscriptaddress,
  showpacs,nofootinbib]{revtex4}
\newcommand{\PRE}[1]{{#1}} 

\usepackage{hyperref}      
\usepackage{bm}
\usepackage{epsfig}

\newcommand{\gweak}{g_{\text{weak}}}
\newcommand{\mweak}{m_{\text{weak}}}
\newcommand{\mplanck}{m_{\text{Pl}}}

\newcommand{\sigmaan}{\sigma_{\text{an}}}

\newcommand{\ev}{\text{eV}}

\newcommand{\mev}{\text{MeV}}
\newcommand{\gev}{\text{GeV}}
\newcommand{\tev}{\text{TeV}}
\newcommand{\pb}{\text{pb}}

\newcommand{\cm}{\text{cm}}

\newcommand{\s}{\text{s}}
\newcommand{\ns}{\text{ns}}

\newcommand{\eqref}[1]{Eq.~(\ref{#1})}
\newcommand{\eqsref}[2]{Eqs.~(\ref{#1}) and (\ref{#2})}
\newcommand{\Eqref}[1]{Equation~(\ref{#1})}
\newcommand{\secref}[1]{Sec.~\ref{sec:#1}}
\newcommand{\secsref}[2]{Secs.~\ref{sec:#1} and \ref{sec:#2}}

\newcommand{\appref}[1]{Appendix~\ref{sec:#1}}

\newcommand{\TBBN}{T_{\text{BBN}}}
\newcommand{\xiBBN}{\xi_{\text{BBN}}}

\newcommand{\mgravitino}{m_{3/2}}

\newcommand{\wino}{\tilde{W}}
\newcommand{\mwino}{m_{\wino}}

\newcommand{\gluino}{\tilde{g}}
\newcommand{\squark}{\tilde{q}}
\newcommand{\gluinoh}{\gluino^h}
\newcommand{\squarkh}{\squark^h}

\begin{document}

\preprint{UCI-TR-2011-02}

\title{ \PRE{\vspace*{1.5in}} 
WIMPless Dark Matter from Non-Abelian
Hidden Sectors with Anomaly-Mediated Supersymmetry Breaking
\PRE{\vspace*{0.3in}} }

\author{Jonathan L.~Feng}
\affiliation{Department of Physics and Astronomy, University of
California, Irvine, CA 92697, USA
\PRE{\vspace*{.2in}}
}

\author{Yael Shadmi\PRE{\vspace*{.5in}}}
\affiliation{Physics Department, Technion-Israel Institute of
Technology, Haifa 32000, Israel
\PRE{\vspace*{.5in}}
}

\date{February 2011}

\begin{abstract}
\PRE{\vspace*{.3in}} In anomaly-mediated supersymmetry breaking (AMSB)
models, superpartner masses are proportional to couplings squared.
Their hidden sectors therefore naturally contain WIMPless dark matter,
particles whose thermal relic abundance is guaranteed to be of the
correct size, even though they are not weakly-interacting massive
particles (WIMPs).  We study viable dark matter candidates in WIMPless
AMSB models with non-Abelian hidden sectors and highlight unusual
possibilities that emerge in even the simplest models.  In one example
with a pure SU($N$) hidden sector, stable hidden gluinos freeze out
with the correct relic density, but have an extremely low, but
natural, confinement scale, providing a framework for self-interacting
dark matter.  In another simple scenario, hidden gluinos freeze out
and decay to visible Winos with the correct relic density, and hidden
glueballs may either be stable, providing a natural framework for
mixed cold-hot dark matter, or may decay, yielding astrophysical
signals.  Last, we present a model with light hidden pions that may be
tested with improved constraints on the number of non-relativistic
degrees of freedom.  All of these scenarios are defined by a small
number of parameters, are consistent with gauge coupling unification,
preserve the beautiful connection between the weak scale and the
observed dark matter relic density, and are natural, with relatively
light visible superpartners.  We conclude with comments on interesting
future directions.
\end{abstract}


\maketitle

\section{Introduction}
\label{sec:introduction}

The thermal relic density of a dark matter candidate $X$ is
\begin{equation}
\Omega_X \propto \frac{1}{\langle \sigmaan v \rangle} \sim
\frac{m_X^2}{g_X^4} \ ,
\end{equation}
where $\langle \sigmaan v \rangle$ is the thermally-averaged
annihilation cross section, and $m_X$ and $g_X$ are the characteristic
mass scale and coupling determining this cross section.  For
weakly-interacting massive particles (WIMPs), the characteristic
values are $m_X \sim \mweak \sim 100~\gev$ and $g_X \sim \gweak \simeq
0.65$, and the thermal relic density is roughly of the desired order
of magnitude, $\Omega_X \sim 0.1$. This coincidence, the WIMP miracle,
is a leading motivation for WIMPs and has guided many searches for
dark matter in particle physics experiments.

At the same time, the relic density constrains only one combination of
$m_X$ and $g_X$.  In the standard model (SM), the only possible value
for $g_X$ is $\gweak$, since dark matter with significant
electromagnetic or strong interactions is essentially excluded.
However, if dark matter is in a hidden sector with its own
interactions, other combinations of $m_X$ and $g_X$ can yield the
correct thermal relic density.  This is the possibility realized in
WIMPless models~\cite{Feng:2008ya,Feng:2008mu}, where dark matter is
hidden, with no SM gauge interactions.  In these models, the dark
matter's mass $m_X$ is not necessarily near $\mweak$, and its hidden
sector gauge couplings $g_X$ are not necessarily near $\gweak$, but
\begin{equation}
\frac{m_X^2}{g_X^4} \sim \frac{\mweak^2}{\gweak^4} \ .
\label{relation}
\end{equation}
WIMPless dark matter particles therefore have the correct thermal
relic density, but with a broad range of possible masses and
couplings.  In addition, their interaction strengths with SM particles
may vary greatly, depending on the presence or absence of connector
particles that induce dark matter-SM interactions through non-gauge
interactions. The WIMPless framework therefore preserves key virtues
of WIMPs, but is far more general, leading to novel implications for
direct~\cite{Feng:2008ya,Feng:2008dz} and indirect dark matter
searches~\cite{Feng:2008qn,Kumar:2009ws,Barger:2010ng,Zhu:2011dz},
precision experiments~\cite{McKeen:2009rm,Yeghiyan:2009xc,%
McKeen:2009ny,Badin:2010uh}, high energy
colliders~\cite{Alwall:2010jc,Goodman:2010ku}, and
cosmology~\cite{Feng:2009mn}.

\Eqref{relation} is required for a thermal relic to match cosmological
observations, but it is also motivated by particle physics
considerations alone.  For example, the new physics flavor and CP
problems motivate supersymmetric (SUSY) models with gauge-mediated
SUSY breaking (GMSB), where generation-blind superpartner masses are
generated by gauge interactions.  The resulting masses are $m_X
\propto g_X^2$.  If the constant of proportionality is similar in both
the visible and hidden sectors, a stable hidden superpartner will
satisfy \eqref{relation} and be an excellent WIMPless dark matter
candidate~\cite{Feng:2008ya,Feng:2008mu}.  Beyond GMSB, however, the
model-building possibilities for WIMPless dark matter have not been
extensively studied.

In this work, we explore possible realizations of the WIMPless miracle
in SUSY models with anomaly-mediated SUSY breaking
(AMSB)~\cite{Randall:1998uk,Giudice:1998xp}.  As in the case of GMSB,
AMSB models are motivated in large part by their potential to solve
the new physics flavor and CP problems through generation-blind
superpartner masses.  These masses are again proportional to couplings
squared, and so AMSB models are also natural homes for WIMPless dark
matter.  In the AMSB framework, one assumes that the minimal
supersymmetric standard model (MSSM) is ``sequestered'' from the
SUSY-breaking sector, that is, it does not have tree-level couplings
to the SUSY-breaking sector.  The visible sector's superpartner masses
are then generated purely by the Weyl anomaly,
\begin{equation}
m_v \sim \frac{g_v^2}{16 \pi^2} \mgravitino \ ,
\label{amsbmassesv}
\end{equation}
where $\mgravitino \sim 100~\tev$ is the gravitino mass.  The same
would hold for any hidden sector of the theory that is similarly
sequestered from SUSY breaking, leading to
\begin{equation}
m_X \sim \frac{g_X^2}{16 \pi^2} \mgravitino \ .
\label{amsbmasses}
\end{equation}
As a result
\begin{equation}
\frac{m_X}{g_X^2} \sim \frac{1}{16 \pi^2} \mgravitino \sim
\frac{m_v}{g_v^2} \ ,
\end{equation}
and, since $m_v \sim \mweak$ and $g_v \sim \gweak$, \eqref{relation}
holds.

In both GMSB and AMSB, the visible sector does not have a good thermal
relic candidate.  In GMSB, SM superpartners decay to the gravitino.
In AMSB, the Wino is typically the lightest supersymmetric particle
(LSP) and is stable, but it typically annihilates too efficiently to
have the correct thermal relic density. The possibility of a hidden
dark matter candidate with the correct thermal relic density is
therefore as welcome in AMSB as in GMSB.  In fact, in several other
aspects, AMSB models are more ideally suited for WIMPless dark matter
than GMSB models.  First, \eqsref{amsbmassesv}{amsbmasses} immediately
imply \eqref{relation}, and so the WIMPless miracle does not require
any ``model building'' to make the constant of proportionality similar
in the visible and hidden sectors.  And second, since $\mgravitino \gg
\mweak$, hidden superpartners cannot decay to gravitinos, and WIMPless
candidates are automatically stable, at least in the absence of
couplings to the MSSM.  Indeed, in some of our examples, the WIMPless
dark matter is stable merely by virtue of spacetime symmetry and gauge
symmetry.

Thanks to these properties, the models we will consider are extremely
simple. The hidden sector is just an SU($N$) gauge theory with some
number $N_F$ of ``quarks'' in the fundamental representation.  The
simplest example is pure SU($N$), where the stable SU($N$) ``gluino''
is WIMPless dark matter, and the theory is completely specified by $N$
and the hidden gauge coupling $g_X$. Even in this simplest example, we
will find the possibility of interesting astrophysical
implications. We will then consider slightly more complicated theories
with $N_F > 0$ flavors and connector particles mediating hidden
sector-visible sector interactions, again with unusual implications
for experiments and observations.  In all cases, however, the WIMPless
miracle naturally preserves the beautiful connection between the weak
scale and the correct dark matter density.  Although we will not
exhaustively explore the phenomenological consequences of these
scenarios here, we will consider several qualitatively different
model-building possibilities, highlight key constraints, and briefly
mention some of the many possible implications for dark matter
properties and the early Universe.

In \secref{spectra} we derive results for AMSB superpartner spectra in
the visible and hidden sectors, and we discuss relic densities and
cosmological constraints in \secref{cosmoconstraints}.  In
Secs.~\ref{sec:noconnectors}--\ref{sec:gbmodel}, we then present a
number of models that satisfy these constraints, but have
qualitatively different features.  We summarize our conclusions and
potentially interesting future directions in \secref{conclusions}.

\section{Superpartner Spectra and LSPs}
\label{sec:spectra}

\subsection{Visible Sector}

In AMSB models the soft SUSY-breaking masses are determined by the
gravitino mass $\mgravitino$ and the values of the dimensionless
couplings of the theory at the SUSY-breaking scale.  These expressions
are well-known~\cite{Randall:1998uk,Giudice:1998xp} and are given in
\appref{amsbappendix} for a general SUSY model.

In the visible sector, which we assume has the low-energy field
content of the MSSM, this implies that the superpartner spectrum is
highly constrained.  For gaugino masses, the $\beta$-function
coefficients are $(b_1, b_2, b_3) = (33/5, 1, -3)$, and so the gaugino
and gravitino mass parameters are in the ratio
\begin{equation}
M_1 : M_2 : M_3 : \mgravitino \simeq 3.31 : 1 : -10.5 : 372 \ .
\label{ratios}
\end{equation}
Because SU(2) is nearly conformal in the MSSM, the Wino is the
lightest MSSM gaugino.  The current bound from LEP2, $\mwino \agt
100~\gev$, implies $\mgravitino \agt 37~\tev$.  In the scalar
superpartner sector, the soft slepton masses squared turn out to be
negative.  There are many solutions to this
problem~\cite{Pomarol:1999ie,Chacko:1999am,Katz:1999uw,Chacko:2000wq,%
Jack:2000cd,Carena:2000ad,Allanach:2000gu,Kaplan:2000jz,%
ArkaniHamed:2000xj}.  Most of these do not modify the gaugino spectrum
and therefore do not affect our discussion here.

As mentioned in \secref{introduction}, the main assumption in AMSB is
that the MSSM is sequestered from the SUSY-breaking sector. One way to
achieve this is in the context of extra dimensions, with the two
sectors localized on different branes and separated by an extra
dimension~\cite{Randall:1998uk}.  If the hidden sector is localized on
the same brane as the MSSM, it is likewise sequestered from the SUSY
breaking.  This scenario allows for the presence of tree-level
couplings of the hidden sector to the MSSM.  Sequestering also results
if the SUSY-breaking sector is near-conformal over some energy
range~\cite{Luty:2001zv}.  In this case, not just the MSSM, but any
other sector of the theory is sequestered from the SUSY breaking, so
no extra assumptions are needed regarding the dark matter hidden sector.

\subsection{Hidden Sector}

In contrast to the visible sector, there is a great deal of
flexibility in defining the hidden sector's field content.  For a
gauge theory with matter, the results of \appref{amsbappendix} imply
that the gaugino and scalar masses have the form
\begin{eqnarray}
m_{1/2} &\sim& b g^2 \frac{1}{16 \pi^2} \mgravitino \\
m_0^2 &\sim& \left( y^4 - y^2 g^2 - b g^4 \right) 
\left( \frac{1}{16\pi^2} \mgravitino \right)^2 \ ,
\label{scalarmass}
\end{eqnarray}
where $g$ and $y$ represent gauge and Yukawa couplings, $b$ represents
one-loop $\beta$-function coefficients, and positive numerical
coefficients have been neglected.  

For a pure gauge theory, the theory must be non-Abelian so that the
gauginos are massive and can potentially be WIMPless dark matter.  For
a gauge theory with matter there are many possibilities.  We will
consider only theories without Yukawa couplings and restrict our
attention to theories without tachyonic scalars.  These are not
necessarily requirements, and there may well be interesting examples
of WIMPless models in the cases we neglect.  But given these
assumptions, \eqref{scalarmass} shows that even with matter, we are
led to consider only non-Abelian gauge groups.  Non-Abelian gauge
groups and strongly interacting dark matter have been explored
previously~\cite{Dover:1979sn,Mohapatra:1997sc,Baer:1998pg,%
Spergel:1999mh,Arvanitaki:2005fa,Banks:2005hc,Kang:2006yd}, but we are
led to this possibility for completely different reasons than those
explored previously.

To be concrete, we focus in this work on hidden sectors that are
SU($N$) gauge theories with $N_F \ge 0$ light flavors of matter in $N
+ \bar{N}$ representations, and no Yukawa couplings.  We will refer to
the hidden gauginos and gauge bosons as gluinos $\gluinoh$ and gluons
$g^h$, and the hidden matter as squarks $\squarkh$ and quarks $q^h$.
In addition, in what follows, $X$ will denote the hidden LSP (hLSP),
and $\alpha_X \equiv g_X(m_X)^2/4\pi$ will denote the hidden
sector's fine structure constant at the scale $m_X$.

Above the hidden gluino and squark mass scale, the one-loop
$\beta$-function coefficient is $b_H = -3N + N_F$.  The gluino and
squark masses are then
\begin{eqnarray}\label{masses}
m_{\gluinoh} &=& (3N-N_F) \, \frac{\alpha_X}{4 \pi} \, \mgravitino \\
m_{\squarkh}^2 &=& (3N-N_F) \, \frac{N^2-1}{N} \, 
\left(\frac{\alpha_X}{4 \pi} \, \mgravitino\right)^2 \ .
\label{squarkmass}
\end{eqnarray} 
We require $N_F < 3N$ so that the supersymmetric theory is
asymptotically-free and the squarks are non-tachyonic.  For $N_F \leq
2N$, the squark is the hLSP, and for $N_F > 2N$ (and, of course, for
$N_F = 0$) the gluino is the hLSP.

In the absence of couplings to the MSSM, the hLSP is stable, because
it is odd under the SU($N$) sector $R$-parity.  In fact, for some
values of $N$, the stability follows just from spacetime symmetry and
gauge symmetry.  A particularly simple case is pure SU($N$) for which
the gluino is clearly stable, since it's the lightest fermion.  More
generally, a gluino hLSP must decay to an odd number of quarks plus
some number of gluons, but for even $N$, an odd number of fundamentals
and anti-fundamentals does not contain the adjoint representation.
Similarly, a squark hLSP cannot decay to quarks and gluons for even
$N$.

Below the scale of the hLSP mass $m_X$, we are left with a
non-supersymmetric SU($N$) gauge theory with $N_F$ flavors.  For $N_F
< N_*$, with $N_* \sim (2.5 - 3)N$, this theory is believed to
confine~\cite{Dietrich:2006cm,Appelquist:1998rb,Miransky:1996pd,%
Cohen:1988sq,Appelquist:1988yc,Poppitz:2009tw,Poppitz:2009uq}.  As
explained above, we need $N_F < 3N$, and so, at least for small values
of $N$, the theory always confines.  The confinement scale is
\begin{equation}
\Lambda \sim m_X \ \text{exp} \! \left(\frac{2\pi}{b_L \alpha_X}
\right) \ ,
\label{Lambdageneral}
\end{equation}
where $b_L = -\frac{11}{3} N + \frac{2}{3} N_F$ is the
$\beta$-function coefficient of the non-supersymmetric theory.  Below
this scale, the quarks and gluons form color-neutral SU($N$)
composites.  Note that we always take $m_X > \Lambda_H$, where
$\Lambda_H$ is the strong coupling scale of the supersymmetric theory;
otherwise, we would need to work directly in the low-energy effective
theory of the SU($N$) composites.

\section{Cosmological Constraints and Relic Densities}
\label{sec:cosmoconstraints}

We begin by outlining various requirements that all models must
satisfy.  This is not a complete list.  In particular, there are
important constraints from structure formation and halo profiles on
self-interactions and charged dark matter, and from Big Bang
nucleosynthesis (BBN) and other observations on scenarios where hidden
sector particles decay to visible ones.  We will discuss these where
relevant when we present concrete models, starting
in~\secref{noconnectors}.

\subsection{Relic Density of Visible LSPs}

Given the assumption that the neutral Wino is the visible sector's
LSP, it is natural to consider it as a dark matter candidate.
Unfortunately, its thermal relic density is typically small, because
it annihilates efficiently through the $S$-wave process $\wino \wino
\to WW$~\cite{Mizuta:1992ja}.  To obtain $\Omega_{\wino} \approx 0.23$
the Winos must be very heavy, with $\mwino \sim
2~\tev$~\cite{Giudice:1998xp}.  This problem is exacerbated in AMSB by
the hierarchy in gaugino masses, as it implies $m_{\tilde{g}} \sim
20~\tev$, which is far above the weak scale and undermines the
motivation of SUSY as a solution to the gauge hierarchy problem.  To
restore Wino dark matter as a possibility, previous attempts have
abandoned the WIMP miracle and explored the possibility that Winos are
produced not by thermal freeze out, but through non-thermal
mechanisms, such as the late decays of moduli~\cite{Moroi:1999zb} or
$Q$-balls~\cite{Fujii:2001xp}, or by thermal freeze out, but in a
non-standard cosmology~\cite{Nihei:2005qx}.

\subsection{Relic Density of Hidden LSPs}
\label{sec:relichLSP}

Whether the hLSP is the gluino or the squark, its annihilation cross
section, just like the visible Wino's, is not helicity suppressed.
Thanks to the WIMPless miracle, the two cross sections are very
roughly comparable, irrespective of the hidden superpartner mass
scale.  However, there are $N$- and $N_F$-dependent factors that may
enhance the hLSP thermal relic density significantly relative to the
Wino case, because SU(2) in the MSSM is nearly conformal and so the
Wino thermal relic density may be thought of as accidentally low.
Keeping track of these factors, we find
\begin{eqnarray}
\frac{\Omega_{\gluinoh}}{\Omega_{\wino}} &\sim& (3N-N_F)^2 \\
\frac{\Omega_{\squarkh}}{\Omega_{\wino}} &\sim& (3N-N_F)^2 \,
\left(\frac{N^2-1}{N}\right)^2 \ .
\label{ratiosquark}
\end{eqnarray}
The $N$- and $N_F$-dependent factors may be large.  For a pure SU(3)
gauge theory, for example, we find enhancements of $\sim 100$,
compensating for the too-low value of $\Omega_{\wino}$.  We will also
consider scenarios in which the hLSP decays to the Wino.  In this
case, the relic abundance will be diluted by $m_X / \mwino$, but there
is still a significant enhancement to the Wino relic density for a
given $\mgravitino$.  Note that the gauge couplings factor out of the
ratio of abundances, but do appear in mass ratios.

What happens to the hLSP relic density after freeze out?  Conventional
dark matter candidates are neutral under preserved gauge symmetries.
In this case, however, the hLSP is charged, leading to new phenomena.
At temperatures $T \agt \Lambda$, the hLSPs may annihilate through
Sommerfeld-enhanced cross sections.  At $T \alt \alpha_X^2 m_X$, they
may also form hLSP-hLSP bound states, which then rapidly leads to hLSP
pair annihilation.  These effects have been analyzed previously in
various contexts~\cite{Baer:1998pg,Hisano:2006nn,Dent:2009bv,%
Zavala:2009mi,Feng:2009mn}.  For the present scenario, they have a
small $\sim {\cal O}(10\%)$ effect on the hLSP relic density,
essentially because both Sommerfeld-enhanced and bound-state catalyzed
annihilation rates are small compared to the Hubble expansion
rate~\cite{Feng:2009mn}.

For $T \alt \Lambda$, however, the hLSPs will hadronize, potentially
enhancing their annihilation~\cite{Dover:1979sn,Mohapatra:1997sc,%
Baer:1998pg,Spergel:1999mh,Arvanitaki:2005fa,Banks:2005hc,Kang:2006yd}.
In particular, the resulting ``$R$-hadrons'' now have $\sim
1/\Lambda^2$-interactions, and pairs of $R$-hadrons can form bound
states, which potentially leads to rapid hLSP-hLSP
annihilation~\cite{Kang:2006yd}.  This annihilation depends
sensitively on the existence of light states with mass below
$\Lambda$, since, for the two hLSPs to annihilate, the bound states of
pairs of $R$-hadrons must lose energy by radiating light particles.
These issues were studied for the case of SM QCD, but their importance
in the context of a general strongly-interacting hidden sector merits
further study.  Note, however, that hadronization effects become
irrelevant if, for example, the hLSP decays to the visible sector
before $T \sim \Lambda$, or if $\Lambda$ is so low that the hidden
gluinos and gluons have never been cold enough to confine.

\subsection{Relic Density of Hidden Quark-Gluon Composites}
\label{sec:composites}

At $T \sim \Lambda$, the hidden sector quarks and gluons form
SU($N$)-gauge invariant composites, including ``mesons,''
``glueballs,'' and ``baryons,'' with masses of order $\Lambda$.  The
relic abundance of these composites is model-dependent, and it is
useful to distinguish between three qualitatively different scenarios:

\begin{itemize}
\setlength{\itemsep}{0pt}\setlength{\parskip}{0pt}\setlength{\parsep}{0pt}

\item[(c1)] The hidden sector contains massless particles, such as
Goldstone bosons or a photon associated with a new U(1). These provide
a thermal bath to allow the SU($N$) composites to annihilate to
sufficiently low densities.  Note that in this case, the composites
have the usual thermal freeze out, so that for $\Lambda \ll m_X$,
their abundances are much smaller than the hLSP abundance.  We will
see an example of this type in \secref{gbmodel}.

\item[(c2)] There are connector fields that efficiently mediate decays
of the {\em unstable} hidden SU($N$) composites to massless particles
in the visible sector.  This is realized in the models of
\secsref{beforebbn}{beforebbnhiggs}.

\item[(c3)] There are no massless fields in the hidden sector and no
efficient decays to the visible sector.  Some SU($N$) composites will
then be stable, either because they are charged under some symmetry,
or because they are the lightest states in the hidden sector.
Requiring that the SU($N$) composites not overclose the Universe then
places an upper bound on $\Lambda$.  For example, consider the
simplest case of pure SU($N$), whose lightest glueballs are stable.
Let the visible sector's temperature be $T$, and assume the hidden
temperature is similar.  For $T \agt \Lambda$ the gluons have thermal
energy density $\rho_{\text{th}} \propto T^4$, at $T \sim \Lambda$ the
gluons form glueballs with mass $\sim \Lambda$, and for $T \alt
\Lambda$, the glueball energy density is $(\Lambda / T)
\rho_{\text{th}} \propto \Lambda T^3$.  The resulting glueball relic
density now is $\Omega \sim \Lambda / 100~\ev$.  Requiring that the
glueballs not have relic density larger than the observed dark matter
density, and, even more stringently, not be too large a contribution
to hot dark matter~\cite{Viel:2005qj,Boyarsky:2008xj} implies $\Lambda
\alt 10~\ev$.  The model of \secref{noconnectors} is an example of
this type.
\end{itemize}

\subsection{Hidden Sector Contributions to $g_*$}
\label{sec:gstar}

Light degrees of freedom contribute to the expansion rate of the
Universe and are constrained by
BBN~\cite{Cyburt:2004yc,Fields:2006ga}.  In the context of hidden
sectors, the current bound from BBN requires~\cite{Feng:2008mu}
\begin{equation}
g_*^h \left[ \frac{\TBBN^h}{\TBBN^v} \right]^4 
\le 2.52 \ (\text{95\% CL}) \ , 
\label{bbnbound}
\end{equation}
where $g_*^h$ is the effective number of non-relativistic degrees of
freedom in the hidden sector at the time of BBN, and $\TBBN^h$ and
$\TBBN^v$ are the temperatures of the hidden and visible sectors at
the time of BBN, respectively.

If $\TBBN^h = \TBBN^v$, the constraint from BBN on $g_*^h$ is
stringent.  For $m_X> \TBBN^h$, the superpartners in the hidden sector
are too heavy to contribute to $g_*^h$.  However, if $\Lambda <
\TBBN^h$, the hidden quarks and gluons contribute $g_*^h = 2 (N^2-1) +
\frac{7}{2} N_F$.  Even in the minimal case with $N=2$ and $N_F = 0$,
this exceeds the bound of \eqref{bbnbound} by more than a factor of 2.

The bound may be evaded in several ways, however, depending on
the confinement scale $\Lambda$:
\begin{itemize}
\setlength{\itemsep}{0pt}\setlength{\parskip}{0pt}\setlength{\parsep}{0pt}

\item[(d1)] $\Lambda\alt \TBBN^h \, (\sim \mev)$.  In this case, the
counting above applies, and to evade the bound, the hidden sector must
be colder than the visible sector at the time of BBN.  If the hidden
sector is completely hidden, it is quite natural for $\TBBN^h$ and
$\TBBN^v$ to be
different~\cite{Kolb:1985bf,Hodges:1993yb,Berezhiani:1995am}.  The
model of \secref{noconnectors} is an example of this type.  If, on the
other hand, there are connector fields coupling the visible and hidden
sectors, the reheat temperature must be below the mass of the
connector fields.  This possibility is realized in the models
of \secsref{stable}{stablehiggs}, where the connectors are very heavy,
and this requirement is not very stringent.  As an added bonus, in
this scenario, the connector fields may be stable, as they are
inflated away and not regenerated after reheating, avoiding
overclosure constraints.

\item[(d2)] $\Lambda \agt \TBBN^h \, (\sim \mev)$. At temperatures
below $\Lambda$ and above $\TBBN^h$, the (unstable) SU($N$) composites
decay to visible sector (MSSM) fields. We will see examples of this
type in \secsref{beforebbn}{beforebbnhiggs}.

\item[(d3)] $\Lambda \agt \TBBN^h \, (\sim \mev)$. At temperatures
below $\Lambda$ and above $\TBBN^h$, the (unstable) SU($N$) composites
decay to light hidden sector fields.  A simple realization of this
possibility is decays to massless Goldstone bosons in the hidden
sector.  We will consider such an example in~\secref{gbmodel}, with
$N_F=2$, so that there are 3 massless scalar Goldstone bosons, which
is marginally consistent with \eqref{bbnbound}.

\end{itemize}

\section{A Pure SU($N$) Hidden Sector without Connectors}
\label{sec:noconnectors}

We begin with a very simple model in which the hidden sector is a pure
SU($N$) gauge theory without matter, and there are no connector fields
coupling the visible and hidden sectors.  The stable hidden gluino is
WIMPless dark matter.  The model is completely specified by
$\mgravitino$, $m_X$, and $N$.  In terms of these, the hidden gauge
coupling is determined by
\begin{equation}
m_X = 3 N \frac{\alpha_X}{4\pi} \mgravitino \ ,
\label{mtilde}
\end{equation}
and the confinement scale is 
\begin{equation}
\Lambda 
\sim m_X \ \text{exp} \! \left( \frac{-6 \, \pi}{11 N \alpha_X} \right) 
= m_X \ \text{exp} \! \left( \frac{-9 \, \mgravitino}{22 m_X} \right)
\simeq m_X \ 10^{-66 \, \mwino/ m_X } \ .
\label{Lambda}
\end{equation}

Because there are no connectors, the hidden gluons $g^h$ (and
glueballs $(g^h g^h)$, if they form) are also stable.  As a result,
the constraint (c3) on the glueball relic density discussed in
\secref{composites} applies, requiring $\Lambda \alt 10~\ev$.  The
hidden sector is therefore weakly-coupled at BBN and contributes
$g_*^h = 2 (N^2 - 1)$ relativistic degrees of freedom at BBN.  The
bound of \eqref{bbnbound} then implies
\begin{equation}
\xiBBN \equiv \frac{\TBBN^h}{\TBBN^v} \le 
\left( \frac{1.26}{N^2 - 1} \right) ^{\frac{1}{4}} \ ;
\end{equation}
although the hidden sector cannot be at the same temperature as the
visible sector, the BBN constraint is satisfied if the hidden sector
is just slightly colder.  Note that, without connectors, it is quite
natural for the visible and hidden sectors to be at different
temperatures.
 
Hidden gluinos annihilate to hidden gluons through $S$-wave processes
with cross section
\begin{equation}
\sigma ( \tilde{g}^h \tilde{g}^h \to g^h g^h ) v \simeq \sigma_0 \ ,
\end{equation}
where 
\begin{equation}
\sigma_0 = k \frac{\pi \alpha_X^2}{m_X^2} \ ,
\end{equation}
and $k$ is an ${\cal O}(1)$ $N$-dependent coefficient.  Using the
results of \appref{relicdensity} for thermal freeze out in a hidden
sector~\cite{Feng:2008mu,Das:2010ts}, the hidden gluino's thermal
relic density is, then,
\begin{equation}
\Omega_X \simeq 0.23 \ \xi_f \, \frac{1}{k}
\left[ \frac{0.025}{\alpha_X} \right]^2
\left[ \frac{m_X}{\tev} \right]^2
\simeq 0.23 \ \xi_f \, \frac{N^2}{k}
\left[ \frac{\mgravitino}{170~\tev} \right]^2 \ ,
\label{omegax}
\end{equation}
where we have used \eqref{mtilde}.  The relic density is independent
of $\alpha_X$ and $m_X$, and is automatically of the right order of
magnitude because the hierarchy problem implies $\mgravitino \sim
100~\tev$; in short, this scenario realizes the WIMPless miracle.  Of
course, although $\Omega_X$ is insensitive to $m_X$ and $\alpha_X$,
the dark matter's properties are not.  In particular, the confinement
scale $\Lambda$ is extremely sensitive to these parameters.

As an example, consider $N=3$.  In this case, $k =
27/64$~\cite{Baer:1998pg}, and $\xiBBN$, the hidden to visible
temperature ratio at BBN, may be as large as 0.63.  Taking this
temperature ratio at freezeout to be $\xi_f = 0.5$, the correct hidden
gluino relic density is achieved for $\mgravitino = 52~\tev$ and
$\mwino = 140~\gev$.  In terms of $m_X$, the coupling is
\begin{equation}
\alpha_X \simeq 0.027 \, \frac{m_X}{\tev} \ ,
\end{equation}
and the confinement scale is
\begin{equation}
\label{lambdamx}
\Lambda \sim m_X \, (5.8 \times 10^{-10})^{\frac{\tev}{m_X}} \ .
\end{equation}

The constraint $\Lambda \alt 10~\ev$ implies $m_X \alt 850~\gev$.
Hidden gluons have a temperature that is roughly similar to the CMB
temperature in the visible sector.  Hidden gluinos have velocity
dispersions that drop to $\sim 10^{-8}$, corresponding to temperatures
$m_X v^2 \sim 10^{-4}~\ev$, at redshifts $z \sim
100$~\cite{Profumo:2006bv,Kamionkowski:2008gj,Feng:2009mn}, before
being sped back up to the current velocity $v \sim 10^{-3}$.  For
$0.1~\ev \alt \Lambda \alt 10~\ev$, then, both the hidden gluons and
gluinos cool to a temperature below $\Lambda$ before redshift $z \sim
100$, and so form $(g^h \gluinoh)$ bound states.  If these remain
intact, these bound states interact through a short-range force with
cross section $\sigma \sim \Lambda^{-2}$.  This violates Bullet
Cluster bounds on dark matter self-interactions, which require $\sigma
/ m_X \alt 3000~\gev^{-3}$~\cite{Markevitch:2003at,Randall:2007ph}.
On the other hand, when the bound states are sped back up to $v \sim
10^{-3}$, collisions may disassociate the bound states, and the
relevant bound is on long-range interactions, as we now discuss.

For $\Lambda \alt 0.1~\ev$ ($m_X \alt 750~\gev)$, there is never a
time at which both hidden gluinos and gluons have a temperature below
$\Lambda$, and so at least some of the hidden gluinos and gluons
remain unbound.  In this case, the result is a hidden gluon and gluino
plasma, and the relevant bounds are not those on short-range
interactions, but those on dark matter interacting through long-range
forces~\cite{Ackerman:2008gi,Feng:2009mn,Feng:2009hw,Buckley:2009in}.
The self-interactions are generically weak enough to avoid constraints
from the Bullet Cluster, and for $m_X \sim 750~\gev$ are marginally
consistent with other bounds, such as those from the observation of
elliptical halos~\cite{Feng:2009mn}.  Further work is required to
determine if such scenarios are truly viable.  
Note, however, that extremely low values of $\Lambda$ occur naturally
in this scenario, and it is remarkable that this first example already
leads to potentially interesting dark matter properties and provides
an extremely simple framework for studying such phenomena.

\section{A Pure SU($N$) Hidden Sector with Connectors to MSSM Gauginos}
\label{sec:connectorsgauginos}

We now consider models with heavy connector fields that mediate
interactions between the hidden and visible sectors.  As in the
previous section, we consider a hidden sector that is pure SU($N$),
that is, without light flavors, so the hidden particle content
consists of just gluons and gluinos, with the gluino mass
of~\eqref{mtilde}.  The gluinos freeze out, but then decay to visible
sector particles through connector-induced higher-dimension operators.
Dark matter will be the conventional MSSM Winos, but, unlike in
standard scenarios, these Winos will inherit their relic density from
hidden gluinos, and this relic density will be naturally in the
correct range because of the WIMPless miracle.

In these scenarios, the hidden gluons may form glueballs, and these,
too, can in principle decay to MSSM fields via loops of connector
fields.  The decay times and final state are determined by the details
of the connector fields.  We will discuss two examples of
connectors. In this section, we consider connectors that give rise to
dimension-8 operators coupling the hidden and visible gauge sectors.
In the next section, we will discuss a larger connector sector that
couples the hidden gauge sector to the MSSM Higgs fields through
dimension-6 operators.

\subsection{Connectors}
\label{sec:connectors}

To preserve the possibility of gauge coupling
unification~\cite{Dimopoulos:1981zb,Dimopoulos:1981yj,%
Sakai:1981gr,Ibanez:1981yh,Einhorn:1981sx}, we introduce connectors in
complete multiplets of the MSSM SU(5) gauge group.  We will add $N_Y$
vector-like connectors $Y$ and $\bar{Y}$ that transform as $(5,N)$ and
$(\bar{5}, \bar{N})$ under SU(5)$\times$ SU($N$)$^h$, respectively,
with a large supersymmetric mass $M_Y$.  This scenario is therefore
specified by 5 fundamental parameters: $\mgravitino$, $m_X$, $N$,
$N_Y$ and $M_Y$.

As we will see below, we will need the hidden gluinos to be
short-lived enough to avoid bounds from BBN.  This can be arranged by
having many light connectors.  What are the bounds on $M_Y$ and
$N_Y$?

For $M_Y$ above $\mgravitino$, the connectors have no effect on the
AMSB soft masses to leading order in the supersymmetry breaking.
Their contributions to the soft masses are therefore suppressed by
$\mgravitino/M_Y$ compared to the AMSB soft masses.  In fact, the size
of these contributions is known, since the connectors behave just like
the messengers of gauge mediation.  We can obtain the connectors'
spectrum by rescaling their superpotential mass term by the
compensator, $M_Y Y \bar{Y} \to M_Y(1+ \mgravitino \theta^2) Y
\bar{Y}$, leading to fermion mass $M_Y$ and scalar masses
\begin{equation}
m_{\tilde{Y}}^2 = M_Y^2\left(1\pm \frac{\mgravitino}{M_Y}\right) \ ,
\end{equation} 
just like GMSB messengers with mass $M_Y$ and a supersymmetry-breaking
parameter $F=M_Y \mgravitino$.  Integrating out the connectors, we get
loop corrections to the soft masses of the visible and hidden sectors.
These are known for arbitrary $F /
M_Y^2$~\cite{Dimopoulos:1996gy,Martin:1996zb}.  The leading order term
in $F/M_Y^2$ cancels the connectors' contributions to the AMSB soft
masses above $M_Y$~\cite{Pomarol:1999ie,Katz:1999uw}.  The
higher-order terms give corrections to the leading-order AMSB soft
masses that are less than $4\%$ even for $M_Y = 2 \mgravitino$, and so
we may take $M_Y$ as light as $2 \mgravitino$ without distorting our
other results.

As for $N_Y$, there is no strict upper bound, but the desire for
perturbativity up to high scales and gauge coupling unification
provides a strong motivation for low $N_Y$.  For $M_Y \sim 100~\tev$,
the requirement that gauge couplings remain perturbative up to the GUT
scale is that the effective number of $5 + \bar{5}$ multiplets
satisfies $N_5 \le 5$. In this case, $N_5 = N N_Y$; given $N \ge 2$,
this implies $N_Y \le 2$.  For larger $M_Y$, this constraint is
weaker.

\subsection{Decay Lifetimes}

Box diagrams with $Y$ particles in the loop mediate decays $\gluinoh
\to g^h g \gluino, g^h W \wino, g^h B \tilde{B}$.  At energies below
$M_Y$, the box diagrams induce the operator
\begin{eqnarray}
\lefteqn{\frac{g_X^2 g_{\text{SM}}^2}{16\pi^2} \frac{2 N_Y}{M_Y^4}
\int d^4\theta \, \bar{W}^h_{\dot{\alpha}} 
\bar{W}^{\dot{\alpha}}  W^{h \, \alpha}  W_\alpha } \nonumber  \\
&& = \alpha_X \alpha_{\text{SM}} \frac{2 N_Y}{M_Y^4} \, 
\left[ \bar{\lambda}^h (\sigma\! \cdot \! \partial) \lambda 
\, F^{h \, \rho\sigma} F_{\rho\sigma}
+ F^h_{\mu\nu} F^{\mu\nu} \, F^{h \, \rho\sigma} F_{\rho\sigma}
\right] ,
\label{operator}
\end{eqnarray}
where the bars stand for complex conjugation, and $g_{\text{SM}}$
stands for the appropriate SM gauge coupling.  {\em A priori}, decays
to all SM gauge bosons are allowed.  In some cases, some of these
decays may be kinematically forbidden.  For example, the constraint
$\Lambda \alt 10~\ev$, together with \eqref{Lambda}, implies $m_X \alt
6 \mwino$, and so decays to MSSM gluinos are not allowed.  In the
following, we will focus on the decay to visible Winos, since this
decay channel is always allowed if any of them are.\footnote{We will
discuss examples in which decays to gluinos are important
in~\secref{beforebbn}.  Decays to Binos are negligible in all our
examples.}

For hidden gluinos and visible Winos that are comparable in mass, but
not particularly degenerate, the decay width to Winos is
\begin{equation}
\Gamma( \gluinoh \to g^h W \wino) \sim
\frac{m_X}{8\pi} \frac{1}{16\pi^2} \, 3 \left(  \alpha_X \alpha_2 \frac{2
  N_Y}{M_Y^4} \right)^2 m_X^8 \ ,
\label{gluinohwidth}
\end{equation}
where $1/(16 \pi^2)$ is the 3-body decay suppression factor, and the
factor of 3 comes from summing over the 3 possible charge combinations
of Winos and $W$ bosons in the final state.  Using $\alpha_2 \sim
1/30$ and \eqref{mtilde}, we find that the hidden gluino lifetime is
\begin{equation}
\tau( \gluinoh \to g^h W \wino) \sim
0.3~\s \ \frac{N^2}{N_Y^2} 
\left[ \frac{\mgravitino}{100 m_X} \right]^{10} 
\left[ \frac{M_Y}{2 \mgravitino} \right]^{8} 
\frac{\tev}{m_X} \ ,
\label{gluinohlifetime}
\end{equation}
where we have normalized $M_Y$ to a fairly low value, as discussed in
\secref{connectors}.

The operator of \eqref{operator} also mediates glueball decay to pairs
of MSSM gauge bosons (see also~\cite{Juknevich:2009ji}).  The dominant
decay is to SM gluons, with a decay width
\begin{equation}
\Gamma( (g^h g^h) \to g g ) \sim
\frac{\Lambda}{8\pi} \, 8 \left(  \alpha_X \alpha_3 \frac{2
  N_Y}{M_Y^4} \right)^2 \Lambda^8 \ ,
\label{glueballwidth}
\end{equation}
implying a lifetime of roughly
\begin{equation}
\tau ( \, (g^h g^h) \to g g ) \sim 
10^{-4} \left[ \frac{m_X}{\Lambda} \right]^9 \tau 
(\gluinoh  \to g^h W \wino ) \ .
\label{glueballlifetime}
\end{equation}
Note that here we have not distinguished between the glueball mass and
$\Lambda$.  Glueball masses in pure glue theories have been calculated
on the lattice and are typically larger than $\Lambda$; for example,
see Ref.~\cite{Morningstar:1999rf} for the case of SU(3), which is a
good example, since we will focus here on small $N$.  The glueball
lifetime of \eqref{glueballlifetime} is extremely sensitive to the
glueball mass, so a more careful treatment of glueball masses would
result in a significantly faster glueball decay than the estimate of
\eqref{glueballlifetime}.  This would make it easier to satisfy the
BBN constraints discussed below, but to be conservative, we will not
include such refined estimates.  Note, however, that a very small
change in $m_X$ or $\alpha_X$ may produce a large change in $\Lambda$
to compensate for such missing factors, and so we expect the
qualitatively distinct possibilities we identify below to remain in
more detailed analyses.

The implications of the lifetime estimates of
\eqsref{gluinohlifetime}{glueballlifetime} may be clarified if we
further require that the Winos from $\gluinoh$ decay have the correct
relic density to be all of dark matter. To implement this constraint,
it will be convenient to define the ratio of hidden gluino to Wino
masses,
\begin{equation}
R \equiv \frac{m_X}{\mwino} \ .
\end{equation}
The Wino relic density is the $\gluinoh$ relic density of
\eqref{omegax} diluted by the ratio of masses, or
\begin{equation}
\Omega_{\wino} \simeq 0.23 \ \xi_f \, \frac{N^2}{k} \frac{1}{R} 
\left[ \frac{\mgravitino}{170~\tev} \right]^2 
\simeq 0.23 \ \xi_f \, \frac{N^2}{k} \frac{1}{R} 
\left[ \frac{\mwino}{460~\gev} \right]^2 \ .
\label{omegawino}
\end{equation}
For $\xi_f \sim k \sim 1$, Winos from hidden gluino decays are all of
the dark matter for
\begin{equation}
\mwino \sim \frac{\sqrt{R}}{N}~500~\gev \ . 
\label{mwinoomega}
\end{equation}
Assuming this, the $\gluinoh$ lifetime is
\begin{equation}
\tau( \gluinoh \to g^h W \wino) \sim 1~\s \ 
\frac{N^3}{N_Y^2}
\left[ \frac{M_Y}{2 \mgravitino} \right]^{8}
\left[ \frac{3.0}{R} \right]^{11.5} \ ,
\label{gluinohlifetime2}
\end{equation}
and the glueball lifetime is
\begin{equation}
\tau ( \, (g^h g^h) \to gg ) \sim 10^{-4} \, 10^{594/R} \,  
\tau( \gluinoh \to g^h W \wino) \ .
\label{glueballlifetime2}
\end{equation}

\subsection{Viable Scenarios}

What are the constraints on the $\gluinoh$ and $(g^h g^h)$ lifetimes?
For the hidden gluino, one might think that it must decay before
temperature $\alpha_X^2 m_X$ to prevent gluino-gluino bound states
from forming, thereby enhancing gluino annihilation and ruining the
WIMPless miracle.  As noted in \secref{relichLSP}, though, this is not
required.  The most stringent constraints are associated with BBN.
The decay $\gluinoh \to g^h W \wino$, followed by $W \to q \bar{q}'$,
produces protons and neutrons, which are very dangerous for
BBN. Hidden gluinos must therefore have lifetimes under $\sim 1~\s$.

For glueballs, there are two possibilities.  If they are effectively
stable, they must not contribute too much to hot dark matter, and so
$\Lambda \alt 10~\ev$.  On the other hand, if they are unstable, their
decays are also subject to constraints from BBN.  These may be avoided
if glueballs decay before 1 s.  Of course, this may be too stringent a
requirement: the constraints depend on whether the glueballs decay to
SM gluons, $W$ bosons or photons, and on the decay time.  There are
clearly many possibilities, leading to different constraints and also
many possible signals.

For simplicity, however, here we consider only the two clearly viable
possibilities in which either $\Lambda \alt 10~\ev$ and hidden gluons
or glueballs are long-lived, or $\Lambda \agt \mev$ and glueballs
decay before BBN.

\subsubsection{Low $\Lambda$: $\Lambda \alt 10~\ev$}
\label{sec:stable}

For $\Lambda \alt 10~\ev$, we need $R \alt 6$, where we have used
\eqref{Lambda}.  We then see immediately from
\eqref{glueballlifetime2} that glueballs are extraordinarily
long-lived in this case.  At the same time, for the hidden gluinos to
decay before BBN, we need $R \agt 3.0$.  Given choices of $N$, $N_Y$
and $M_Y$, and assuming the correct Wino relic density, there is then
a one-parameter family of viable models parametrized by $R$ in the
range $3 \alt R \alt 6$.

As an example, consider $N =3$, $N_Y = 1$, and $M_Y = 2 \mgravitino$.
For $R = 5.5$, we find $\mwino \sim 400~\gev$, $m_X \sim 2~\tev$,
$\alpha_X \sim 0.02$, and $\Lambda \sim \ev$.  Assuming $\xi_f \sim k
\sim 1$, hidden gluinos freeze out with $\Omega_X \approx 1$, and then
decay at 0.02~s to MSSM Winos with the right relic density to be all
of dark matter.  Because the hidden gluinos decay early, constraints
on dark matter self-interactions do not apply.  

Alternatively, taking $R = 4$ and $N = N_Y = 2$, we find very similar
values for the masses of the Wino and the hidden gluino, but the
confinement scale is much smaller, with $\Lambda\sim 10^{-5}~\ev$.
The hidden gluinos decay to Winos at 0.07~s, but the hidden gluons
remain unbound and constitute a negligible fraction the Universe's
energy density.

Note that the value of $\Lambda$ may vary widely.  For $\Lambda$ near
its upper bound, these scenarios predict mixed hot-cold dark matter,
with observable implications for small-scale structure.  Note also
that some connectors are stable, as there are no gauge-invariant
decays, so that the reheat temperature must be below $M_Y$.  In this
scenario, however, this is not a very stringent constraint, as the
connectors are very heavy, with $M_Y \sim 100~\tev$, and it is already
motivated by the BBN constraint on $g_*^h$.

\subsubsection{Hidden Glueballs Decaying Before BBN}
\label{sec:beforebbn}

For the glueballs to decay before BBN, we need $\tau ( \, (g^h g^h)
\to gg ) \alt 1~\s$.  At the same time, the hidden gluino must decay
after Wino freeze out at $t \sim 10^{-10}~\s$.  As we will see, these
requirements imply a large $R$, for which the hidden gluino can decay
to the visible gluino.  We therefore require
\begin{equation}
\frac{\tau ( \, (g^h g^h) \to gg )}
{\tau( \gluinoh \to g^h g \tilde{g})}
\sim 30 \cdot 10^{-4}\, \cdot\, 10^{594/R} < 10^{10} 
\Longrightarrow R \agt 45 \ ,
\end{equation}
where the factor of $30 \sim 8 \alpha_3^2 / (3 \alpha_2^2)$ arises
from the enhancement of the decay width to MSSM gluinos over the decay
to MSSM Winos.  As an example, consider $N =6$, and $R = 55$, for
which $\mwino \sim 600~\gev$, $m_X \sim 30~\tev$, $\alpha_X \sim 0.1$,
$\mgravitino \sim 200~\tev$, and $\Lambda \sim 2~\tev$.  The hidden
gluino mass, $m_X$, is quite large, but it is below the unitarity
bound for thermal relics~\cite{Griest:1989wd}.  Taking one set of
connector fields, $N_Y=1$, at $M_Y = 10 \mgravitino$,\footnote{Note
that for $N=6$, the connectors constitute six additional flavors of
the visible SU(3), which still gives a perturbative coupling at the
GUT scale for $M_Y=2000~\tev$.} one finds that the hidden gluino
decays at $t \sim 10^{-8}~\s$, and the glueball decays at $t \sim
1~\s$, avoiding BBN constraints.  The Wino thermal relic density is
negligible, but non-thermal production from hidden gluino decays gives
it the desired relic density.

Note that the hidden gluino decay occurs at temperatures somewhat
below $\Lambda$, and so after the hidden gluino freezes out, it
hadronizes and forms hidden $R$-hadrons before it decays to the Wino.
In principle, this could lead to renewed hidden gluino annihilations,
since the cross section for $R$-hadron interactions is now raised to
$\sim 1/\Lambda^2$.  For these annihilations to occur, the $R$-hadrons
must first form bound states, and later lose energy so that the hidden
gluino pair in the $R$-hadron bound state can actually annihilate;
see, for example, the discussion in Ref.~\cite{Kang:2006yd}.  Both of
these processes require the emission of light particles, which carry
away the binding energy and the energy released when the initial
excited $R$-hadron bound state relaxes to the ground state.  These
energies are characterized by two quantities: $\Lambda$ and
$\alpha_X^2 m_X$.  In the scenarios given here, however, the lightest
particles in the hidden sector are the glueballs, with masses $\agt
\Lambda$, and $\alpha_X^2 m_X<\Lambda$.  Therefore, the hidden gluinos
cannot annihilate effectively even after they hadronize, and they
survive in $R$-hadrons until they decay to Winos.

As in the previous case, this scenario has implications for
observations.  We expect that glueball decay times below 1~s are
allowed, but for lifetimes near this upper bound, these scenarios
predict astrophysical signals, in BBN or other observables sensitive
to late decays.  Finally, note that, also as in the previous case, the
connector fields are stable, and the reheat temperature must again be
below $M_Y \sim 1000~\tev$.  However, $g_*^h=0$ at BBN in this case,
since the hidden glueballs decay to SM fields before BBN.

\section{A Pure SU($N$) Hidden Sector with Connectors to MSSM Higgsinos}
\label{sec:connectorshiggsinos}

\subsection{Connectors}
\label{sec:connectors2}

We now consider an alternative scenario in which the hidden gluinos
decay not to SM gauge bosons, but to SM Higgs bosons.  We add one copy
($N_Y = 1$) of the same connector fields as before, as well as a
vector-like SU($N$) pair $Q$ and $\bar{Q}$, which are singlets under
the SM, so that, in all, the new heavy fields and their
representations under SU(5)$\times$ SU($N$)$^h$ are
\begin{equation}
Y \, (5, N) \ , \quad \bar{Y} \, ( \bar{5}, \bar{N})\ , \quad 
Q \, (1, N) \ , \quad \bar{Q} \, (1, \bar{N}) \ .
\label{conb}
\end{equation}
We couple the $Q$ and the (SU(2) doublets of the) $Y$ connectors to
the MSSM through the superpotential\footnote{Note that once the MSSM
Higgs bosons develop vacuum expectation values, the first two terms in
the superpotential contribute to the connectors' masses, but these
corrections are negligible.}
\begin{equation}
W= y Y\bar{Q} H_d + \bar{y}\bar{Y} Q H_u 
+ M_Y Y\bar{Y} + M_Q Q\bar{Q} \ ,
\end{equation}
where $H_u$ and $H_d$ are the MSSM Higgs supermultiplets.  For
simplicity, we set $M_Y = M_Q \equiv M$ and $\bar{y} = y$.  As in
\secref{connectorsgauginos}, we expect $M \agt 2\mgravitino$ to be
acceptable.  The connector sector is effectively $N$ pairs of
$5+\bar5$ (and 6 SU($N$) flavors), and so gauge coupling unification
is preserved for $N \le 5$.

\subsection{Decay Lifetimes}

As in \secref{connectorsgauginos}, the connectors induce $\gluinoh$
decay through a box diagram, this time with $Q$ and $Y$ connectors in
the loop.  Integrating out the connector fields yields the operator
\begin{eqnarray}
\lefteqn{ 
\frac{g_X^2 y^2}{16\pi^2} \frac{2}{M^2}
\int d^2\theta W^{h\, \alpha} W^h_\alpha H_u H_d } \nonumber \\
&& = \frac{2 \alpha_X \alpha_y} {M^2} \, 
\left(\bar{\lambda}^h \sigma^\mu \bar{\sigma}^\nu \tilde{H_d} 
F^h_{\mu\nu} H_u + F^h_{\mu\nu} F^{h\, \mu\nu} H_u H_d
\right) \ .
\label{higgsoperator}
\end{eqnarray}

The decay width is roughly
\begin{equation}
\Gamma (\gluinoh \to g^h H_u \tilde{H}_d ) 
\sim \frac{m_X}{8 \pi} \frac{1}{16 \pi^2} \, 2 \,
\left(\alpha_X \alpha_y \frac{2}{M^2}
\right)^2 m_X^4 \ ,
\end{equation}
where the loop factor is as in \eqref{gluinohwidth}, and the factor of
2 accounts for the 2 possible charge assignments for the Higgs boson
and Higgsino in the final state.  Using \eqref{mtilde}, the $\gluinoh$
lifetime is, then,
\begin{equation}
\tau (\gluinoh \to g^h H_u \tilde{H}_d) 
\sim 1 \times 10^{-8}~\s \ 
\left[ \frac{N}{2} \right]^2
\left[ \frac{0.01}{\alpha_y} \right]^2
\left[ \frac{\mgravitino}{100 m_X} \right]^6
\left[ \frac{M}{2 \mgravitino} \right]^4
\frac{\tev}{m_X}  \ .
\label{tauhiggs}
\end{equation}
The operator of \eqref{higgsoperator} also mediates glueball decay to
two Higgs bosons.  If kinematically accessible, the glueball decay
width is
\begin{equation}
\Gamma ( \, (g^h g^h) \to H_u H_d ) 
\sim \frac{\Lambda}{8 \pi} \, 2 \,
\left(\alpha_X \alpha_y \frac{2}{M^2} \right)^2 \Lambda^4 \ ,
\end{equation}
corresponding to a lifetime of
\begin{equation}
\tau ( \, (g^h g^h) \to H_u H_d ) \sim 10^{-2}
\left[ \frac{m_X}{\Lambda} \right]^5 \tau ( \gluinoh \to g^h H_u
\tilde{H}_d ) \ ,
\end{equation}
subject to the same uncertainties discussed below
\eqref{glueballlifetime}.

As in \secref{connectorsgauginos}, we may include the constraint from
the relic density.  The relic density is again diluted by the hidden
gluino decay to Winos, and so \eqref{omegawino} again applies.  Using
\eqref{mwinoomega}, we find
\begin{equation}
\tau (\gluinoh \to g^h H_u \tilde{H}_d) 
\sim 1 \times 10^{-4}~\s \ 
\left[ \frac{N}{2} \right]^3
\left[ \frac{0.01}{\alpha_y} \right]^2
\left[ \frac{M}{2 \mgravitino} \right]^4
\frac{1}{R^{7.5}}  \ ,
\label{tauhiggs2}
\end{equation}
and the glueball lifetime satisfies
\begin{equation}
\tau (\, (g^h g^h) \to H_u H_d ) \sim  10^{-2} \, 10^{330/R} \,  
\tau ( \gluinoh \to g^h H_u \tilde{H}_d ) \ .
\end{equation}

\subsection{Viable Scenarios}

We may again identify two qualitatively different classes of viable
scenarios, depending on whether the hidden gluons are effectively
stable, or whether they form glueballs and decay before BBN.  In
contrast to \secref{connectorsgauginos}, however, where the operator
of \eqref{operator} was dimension 8, here the operator of
\eqref{higgsoperator} is only dimension 6.  It is therefore easy to
arrange for very small lifetimes, and the discrepancy between the
gluino and glueball lifetimes is reduced.

\subsubsection{Low $\Lambda$: $\Lambda \alt 10~\ev$}
\label{sec:stablehiggs}

If glueballs are effectively stable, we need $\Lambda \alt 10~\ev$ and
$R \alt 6$.  There are many possible choices of parameters that are
viable.  For example, let $N = 2$, $\alpha_y = 0.01$, and $R=4$.  The
hidden gluino decays may be anywhere in the desired range $1~\ns \alt
t \alt 1~\s$ for $M$ in the range $2\mgravitino \alt M \alt 350
\mgravitino$.  Hidden gluons are very long-lived. The other parameters
are as in \secref{stable}: $\mwino = 400~\gev$, $m_X \sim 2~\tev$,
$\alpha_X \sim 0.02$, and $\Lambda \sim \ev$.  The hidden gluino
freezes out with $\Omega_X \simeq 1$, and then decays to MSSM Winos
with the right relic density to be all of dark matter.  
For $\Lambda$ near its upper bound, this scenario provides a very
simple framework for mixed dark matter, with both MSSM Wino cold and
hidden glueball hot components.

The hidden sector gluons contribute $g_*^h=6$ at BBN, and so the
temperatures of the two sectors must be somewhat different. This is
also motivated by the fact that some connectors are stable, and the
reheat temperature must be below their mass.

\subsubsection{Hidden Glueballs Decaying Before BBN}
\label{sec:beforebbnhiggs}

For the ratio of glueball lifetime to $\gluinoh$ lifetime not to
exceed 10 orders of magnitude,
\begin{equation}
\frac{\tau ( \, (g^h g^h) \to gg )}{\tau( \gluinoh \to g^h g \tilde{g})}
\sim 10^{-2}\, 10^{330/R} < 10^{10} \Longrightarrow R \agt 27 \ .
\end{equation}
Taking, for example, $N=2$, $\alpha_y = 0.01$, $R=30$, and $M = 37
\mgravitino$, we find that the hidden gluino lifetime is around 1 ns,
the hidden glueball decays around 1 s, $\mwino=1.3~\tev$, $m_X \simeq
40~\tev$, $\alpha_X \sim 0.2$, and $\Lambda \sim 200~\gev$. 
The glueball decays may have observable effects in BBN or other
astrophysical signals.

Once again the $\gluinoh$ decays at temperatures a bit lower than
$\Lambda$, but its abundance is not significantly diluted by hadronic
effects.  The hidden sector does not contribute to $g_*$ at BBN, and
so the visible and hidden sectors may be at the same temperature, but
the reheat temperature must be below the mass of the stable connectors
$M_Y \sim 10^3~\tev$.

\section{An SU($N$) Hidden Sector with Flavor and Light Goldstone Bosons}
\label{sec:gbmodel}

So far we have studied three kinds of models.  In one
(\secref{noconnectors}), the hidden sector does not interact with the
visible sector, so that both the hidden dark matter and the hidden
composites are stable.  In the second (\secref{beforebbn},
\secref{beforebbnhiggs}), the hidden dark matter candidate decays to
the Wino, but the hidden SU($N$) composites (glueballs, for the case
of pure Yang-Mills) decay to the visible sector.  In the third
(\secref{stable}, \secref{stablehiggs}) the hidden dark matter
candidate again decays to the Wino, but the hidden composites are
effectively stable.  In this latter case, there is a stringent bound
on the confinement scale $\Lambda \alt 10~\ev$, since otherwise the
hidden glueballs overclose the universe or contribute too much to hot
dark matter.

Here we will consider a qualitatively different example, in which the
hidden sector contains light Goldstone bosons, with masses
significantly below $\Lambda$. The hidden dark matter candidate will
decay to the Wino through loops of connector fields, and the glueballs
will decay to the hidden Goldstone bosons.  Furthermore, the light
Goldstone bosons provide a thermal bath for any stable SU($N$)
composites, such as baryons, so that the resulting relic abundance of
these composites is negligible.

For concreteness, we will focus on the simplest possibility, $N=3$ and
$N_F = 2$, that is, hidden SU(3) with two massless flavors.  
Chiral symmetry breaking results in three Goldstone bosons, which is
marginally consistent with BBN constraints on $g_*^h$, and is testable
with future improvements of these constraints.  The SU(3) confinement
scale is above an MeV, so that the only new light particles at BBN are
the Goldstone bosons.  We will also include connector fields so that
the hLSP decays to the Wino shortly after Wino freezeout.  As we will
see, the connector fields in this example are not stable, so that the
hidden and visible sectors can be in thermal equilibrium.

Because $N_F \le 2 N$, the hLSP is now the hidden squark.
\Eqref{squarkmass} implies that its mass is
\begin{equation}
m_X \simeq 0.34 \, \alpha_X \, \mgravitino \ ,
\end{equation}
and using also \eqref{Lambdageneral}, we find
\begin{equation}
\Lambda \sim 10^{-36/R}\, m_X\ . 
\end{equation}
To get the correct Wino relic abundance from hidden squark decays, we
need
\begin{equation}
\mwino \sim \sqrt{R}\ 300~\gev \ ,
\end{equation}
so we can rewrite $\Lambda$ as
\begin{equation}
\Lambda \sim 300~\gev \, R^{3/2} \, 10^{-36/R} \ .
\end{equation}
Requiring $\Lambda \agt \mev$, we find $R \agt 5$.

We will now add connector fields to the theory, so that the hidden
squark eventually decays to the Wino. As above, we take the connector
fields to be vector-like pairs transforming as bifundamentals under
SU(5)$\times$ SU(3)$^h$:
\begin{equation}
Y \, (5, 3) \ , \quad \bar{Y} \, ( \bar{5}, \bar{3}) \ .
\label{conb2}
\end{equation}
We will need two such pairs, with the superpotential
\begin{equation}
W= y Y^d_i\bar{q}^h_i H_d + 
\bar{y}\bar{Y}^d_i q^h_i H_u + M_Y Y\bar{Y} \ .
\end{equation}
Here $i=1,2$, $q^h$, $\bar{q}^h$ are the hidden SU(3) quarks, and the
superscript $d$ on the $Y$ fields denotes the doublet fields of the
$5$ and $\bar{5}$.  Note that the connectors are unstable: the doublet
$Y$ fields can decay to Higgs fields and hidden quark fields. Since
running effects create a splitting between the doublet and triplet $Y$
fields (see, for example, Ref.~\cite{Dimopoulos:1996gy}), the triplets
can decay to the doublets.

For simplicity, we will set $y = \bar{y}$.  Integrating out the
connector fields we have the following superpotential coupling of the
hidden quarks to Higgs fields:
\begin{equation}
\label{op}
\frac{y^2}{M_Y} q^h_i \bar{q}^h_i H_u H_d \ ,
\end{equation}
which induces hidden squark decay into a hidden quark, Higgs and
Higgsino with lifetime
\begin{equation}
\tau(\tilde{q}^h \to q^h H_u \tilde{H}_d) \sim
3 \times 10^{-24}~\s \ \left[\frac{M_Y}{y^2 m_X}\right]^2 
\, \left[\frac{\tev}{m_X}\right]  
\simeq 1 \times 10^{-18}~\s \ R^{-3.5}\, 
\left[\frac{M_Y}{y^2 \mgravitino}\right]^2 .
\end{equation}
Thus for example, for $R=5$, we can have the hidden squark decay at
$10^{-6}~\s$ for $M_Y= 10^7 y^2 \mgravitino$.

Note that the operator of \eqref{op} induces a small Goldstone boson
mass
\begin{equation}
m_\pi \sim y^2 \frac{\langle H_u H_d \rangle}{M_Y} \ .
\end{equation} 
As discussed in (c3), such masses are constrained by the bound on the
amount of hot dark matter in the Universe.  For $M_Y = 10^7 y^2
\mgravitino$, $m_\pi \sim 10~\ev$, which is consistent with these
bounds.

\section{Conclusions}
\label{sec:conclusions}

Supersymmetric extensions of the SM contain a fundamental mass scale,
the supersymmetry breaking scale, which enters the masses of
superpartners in the visible sector as well as in any hidden
sector. Furthermore, if a hidden sector is truly hidden, with no
interactions with the SM, it generically contains a stable
superpartner, which is protected by the $R$-parity of the hidden
sector.  These two features allow for the construction of dark matter
models in which the dark matter relic abundance is related to the weak
scale.  In AMSB models, this abundance is actually the same as the
usual WIMP abundance, since the dark matter mass is proportional to
its coupling squared, and only their ratio enters in the abundance.
These models thus offer a particularly simple realization of the
WIMPless dark matter idea.

In this paper, we studied dark matter candidates from non-Abelian
hidden sectors with AMSB.  The hidden sectors we consider are very
simple. They are SU($N$) gauge theories, with either no matter or a
few fundamental flavors.  In some of our examples, the hidden LSP is
stable simply as a result of gauge symmetry and supersymmetry, and its
relic abundance is automatically of the correct size by the WIMPless
miracle.  In other examples, the hidden and visible sectors interact
through higher dimension operators, so that the hidden LSP freezes out
and then decays to a visible Wino.  The result is Wino dark matter
which, despite its large annihilation cross section, has the correct
abundance, with favorable implications for indirect detection.

As we have seen, the phenomenology of these models is very rich, owing
partly to the non-Abelian interactions of the dark matter candidate.
As an example, some of these models have a confinement scale $\Lambda$
that is naturally very small, as a result of renormalization group
evolution, with a wealth of potentially interesting astrophysical
implications.  In the model of \secref{noconnectors}, the hidden LSP
is the dark matter, and cannot be seen in any direct or indirect
detection experiment. However, the confinement scale is very small,
and the dark matter is self-interacting through a long-range
non-Abelian force.  In the examples of
\secsref{connectorsgauginos}{connectorshiggsinos}, hidden gluinos
freeze out and decay to visible Winos with the correct relic density.
The accompanying hidden glueballs may either be stable, as discussed
in \secsref{stable}{stablehiggs}, providing a natural framework for
mixed cold-hot dark matter, or may decay, as discussed in
\secsref{beforebbn}{beforebbnhiggs}, yielding astrophysical signals.
We have also presented in \secref{gbmodel} a model with 3 light hidden
pions that contribute to the number of non-relativistic degrees of
freedom at BBN, and will be excluded or favored as constraints on this
quantity improve.  In all of these cases, the scenarios are defined by
a small number of parameters, are consistent with gauge coupling
unification, preserve the beautiful connection between the weak scale
and the observed dark matter relic density, and are natural, with
relatively light visible superpartners.

We have only outlined the main features of representative models here,
and it would be interesting to explore specific models in more detail.
The cosmology of (meta)stable particles with non-Abelian interactions
was studied to some extent for the case of QCD, but even that case has
many unsettled issues.  It would also be interesting to study Abelian
hidden sectors, or hidden sectors with no gauge interactions, but with
Yukawa interactions.  Such hidden sectors are theoretically less
clean, because some model building is required to guarantee the
stability of the hidden LSP, but their phenomenology is likely to be
simpler.

Finally, the models we studied are very predictive, since, because the
superpartner masses are determined by anomaly mediation, they depend
on a very small number of parameters.  They thus offer a particularly
clean realization of WIMPless dark matter.  It would also be
interesting to generalize this idea to other frameworks of
supersymmetry breaking, in which the hidden dark matter abundance does
exhibit some dependence on the hidden dark matter mass and coupling,
but is still related to the weak scale because of the underlying
supersymmetry breaking scale.

\section*{Acknowledgments}

We thank Masahiro Ibe, Manoj Kaplinghat, and Hai-Bo Yu for helpful
conversations.  Much of this work was done during a sabbatical year of
YS at UC Irvine. YS thanks the UC Irvine particle theory group for its
warm hospitality and financial support during this year.  The work of
JLF was supported in part by NSF grants PHY--0653656 and
PHY--0970173. The work of YS was supported by the Israel Science
Foundation (ISF) under Grant No.~1155/07. This research was supported
in part by the United States-Israel Binational Science Foundation
(BSF) under Grant No.~2006071.

\appendix

\section{AMSB Superpartner Masses}
\label{sec:amsbappendix}

In AMSB, the soft SUSY-breaking parameters are determined by the
gravitino mass $\mgravitino$ and the (weak-scale values of the)
dimensionless couplings of the
theory~\cite{Randall:1998uk,Giudice:1998xp}.  Consider a
supersymmetric model with gauge group $G$, gauge coupling $g$ and
Yukawa couplings $y^{ijk}$ defined by the superpotential
\begin{equation}
W = \frac{1}{6} y^{ijk} X_i X_j X_k \ ,
\end{equation}
where the $X_i$ are chiral superfields.  The gauge and Yukawa
coupling renormalization group equations are
\begin{eqnarray}
\dot{g} &=& \frac{1}{16\pi^2} b g^3 \\
\dot{y}^{ijk} &=& y^{pjk} \gamma_p^i
+ y^{ipk} \gamma_p^j + y^{ijp} \gamma_p^k \ ,
\end{eqnarray}
where $b = - 3 C(G) + \sum_i C(i)$, $\dot{(\ )} \equiv d/d\ln
(\mu/Q)$, and
\begin{equation}
\gamma_i^j = \frac{1}{16 \pi^2} 
\left[ \frac{1}{2} y_{imn} y^{jmn} - 2 \delta_i^j g^2 C(i) \right] .
\end{equation}
The group theoretic constants are defined by
\begin{eqnarray}
t^a t^a &=& C(G) \, \bf{1} \\
\text{Tr} \, t^a t^b &=& C(i) \, \delta^{ab}  \ ,
\end{eqnarray}
where the matrices $t^a$ are the generators for representation $i$.
Note that in our conventions, asymptotically-free theories have $b <
0$.

Defining soft SUSY-breaking terms
\begin{equation}
{\cal L}_{\text{soft}} 
= \left\{ - \frac{1}{2} M_{\lambda} \lambda \lambda
- \frac{1}{2} ( m^2 )_i^j \phi^{* \, i} \phi_j 
- \frac{1}{6} A^{ijk} \phi_i \phi_j \phi_k 
+ \text{H.c.} \right\} \ , 
\end{equation}
the AMSB soft SUSY-breaking parameters are
\begin{eqnarray}
M_{\lambda} &=& \frac{1}{16 \pi^2} b g^2 \mgravitino \nonumber \\
(m^2)_i^j &=& \frac{1}{2} \dot{\gamma}_i^j \mgravitino^2 
\label{generalamsb} \\
A^{ijk} &=& - \left( y^{pjk} \gamma_p^i
+ y^{ipk} \gamma_p^j + y^{ijp} \gamma_p^k \right) \mgravitino \ . 
\nonumber
\end{eqnarray}

\section{Thermal Relic Density in a Hidden Sector with a Different 
Temperature}
\label{sec:relicdensity}

Thermal freeze out is modified if it occurs in a sector with a
different temperature from the observable
sector's~\cite{Feng:2008mu,Das:2010ts}.  Here we summarize the main
results.

Assume that a particle $X$ with mass $m_X$ annihilates through
$S$-wave processes with cross section
\begin{equation}
\sigma (XX \to \text{anything} \, ) \, v \approx \sigma_0 \ .
\end{equation}
The particle then freezes out when the hidden and visible sector
temperatures are $T_f^h$ and $T_f^v$, respectively.  The resulting
thermal relic density is
\begin{equation}
\Omega_X \approx \frac{s_0}{\rho_c} 
\frac{3.79 \, x_f} {\left( g_{*S}/\sqrt{g^{\text{tot}}_*} \, \right)
  \mplanck \, \sigma_0} \ ,
\end{equation}
where $s_0 \simeq 2970~\cm^{-3}$ is the visible sector's entropy
density now, $\rho_c \simeq 0.527 \times 10^{4}~\ev~\cm^{-3}$ is the
critical density, $x_f \equiv m_X / T_f^v$, $g_{*S} \sim 100$ and
$g^{\text{tot}}_* \sim 100$ are the visible and total number of
relativistic degrees of freedom at freeze out, and $\mplanck \simeq
1.2~\times 10^{19}~\gev$ is the Planck mass.  The freeze out
temperature is given by
\begin{equation}
x_f = \xi_f \ln L 
- \frac{1}{2} \, \xi_f \ln \left( \xi_f \ln L \right) \ ,
\end{equation}
where
\begin{equation}
\xi_f \equiv \frac{T_f^h}{T_f^v} \ ,
\end{equation}
and 
\begin{equation}
L \approx 0.038 \, \mplanck \, m_X \, \sigma_0 
(g/\sqrt{g^{\text{tot}}_*}) \, \xi^{3/2} \, \delta(\delta+2) \ ,
\end{equation}
where $g$ is the number of $X$ degrees of freedom, and the parameter
$\delta$ is tuned to make these analytical results fit the numerical
results.  For $\xi \sim 0.3 - 1$, $\delta \sim 0.2 - 0.5$ gives a good
fit~\cite{Feng:2008mu}.  

As is well-known, $\Omega_X$ is inversely proportional to $\sigma_0$
and only logarithmically sensitive to $m_X$.  Note, however, that
$\sigma_0$ is also only logarithmically sensitive to $g$.  For
example, for the case where $X$ is a gluino of hidden SU($N$), the
thermal relic density is not enhanced by $N^2 -1$, as it would be for
$N^2-1$ independent degrees of freedom, because the $N^2-1$ gluino
degrees of freedom interact with each other.  As a result, for a wide
range of parameters, $x_f \approx 25 \xi_f$ to a good approximation.

We then find that the thermal relic density is
\begin{equation}
\Omega_X \approx \xi_f \, \frac{0.17~\pb}{\sigma_0} 
\simeq \xi_f \, \frac{1}{\sigma_0} 
\left[ \frac{0.021}{\tev} \right]^2
\simeq 0.23 \ \xi_f \, \frac{1}{k} 
\left[ \frac{0.025}{\alpha_X} \frac{m_X}{\tev} \right]^2 \ ,
\end{equation}
where in the last step, we've parametrized the cross section as
$\sigma_0 = k \pi \alpha_X^2 / m_X^2$.  The final result is,
therefore, simple: for a thermal relic that freezes out in a hidden
sector with a different temperature, the thermal relic density is
modified by the factor $\xi_f \equiv T_f^h/T_f^v$ from the standard
result.


\providecommand{\href}[2]{#2}\begingroup\raggedright\endgroup

\end{document}